# Dirac's hole theory versus field theory for a time dependent perturbation

by

Dan Solomon

Rauland-Borg Corporation
3450 W Oakton
Skokie, IL 60076
USA

Phone – 1-847-324-8337
Email: dan.solomon@rauland.com

November 23, 2003

## Abstract

Dirac's hole theory (HT) and quantum field theory (QFT) are generally considered to be equivalent to each other. However, it has been recently shown that for a time independent perturbation different results are obtained when the change in the vacuum energy is calculated. Here we shall extend this discussion to include a time dependent perturbation. It will be show that HT and QFT yield different results for the change in the vacuum energy due to a time dependent perturbation.

I. Introduction

Dirac's hole theory and quantum field theory are generally assumed to be equivalent. Recently several papers have appeared in the literature pointing out that there are differences between Dirac's hole theory (HT) and quantum field theory (QFT) [1][2][3][4]. The problem was originally examined by Coutinho et al[1][2]. They calculated the second order change in the energy of the vacuum state due to a time independent perturbation. They found that HT and QFT yielded different results. They concluded that the difference between HT and QFT was related to the validity of Feynman's belief that the Pauli Exclusion principle can be disregarded for intermediate states in perturbation theory. This belief was based on Feynman's observation that terms that violate the Pauli principle formally cancel out in perturbation theory. However Coutino et al show that this is not necessarily the case for HT when applied to an actual problem. This author (Solomon [4]) found that this problem was related to the way the vacuum state was defined in QFT. If the definition of the vacuum state was modified as described in [4] then the HT and QFT would yield identical results.

The previous work examined this problem for the case of a time independent perturbation. In this article the differences between HT and QFT for a time dependent perturbation will be examined. In Section II we derive a formal expression for the quantity $\Delta E_{hvac}^{(2)}$ which is the second order change in the vacuum energy in HT. In Section III we derive an expression for $\Delta E_{qvac}^{(2)}$ which is the second order change in the vacuum energy in QFT. We show that these expressions are different due to the fact that $\Delta E_{hvac}^{(2)}$ includes a quantity called $X_{2,N}(N \to \infty)$ which is not included in $\Delta E_{qvac}^{(2)}$. In order, then, for $\Delta E_{hvac}^{(2)}$ and $\Delta E_{qvac}^{(2)}$ to be equal $X_{2,N}(N \to \infty)$ must be zero for all possible perturbations. In order to test this possibility an actual problem is worked out in Section IV. Here it is shown that for this particular problem $\Delta E_{hvac}^{(2)} < 0$ and $\Delta E_{qvac}^{(2)} = 0$. Therefore $X_{2,N}(N \to \infty) \ne 0$ and $\Delta E_{hvac}^{(2)} \ne \Delta E_{qvac}^{(2)}$. This result is confirmed in Section V where the value of $X_{2,N}(N \to \infty)$ is actually calculated and is shown to be nonzero. It is then shown in Section VI that by modifying the definition of the QFT vacuum state the difference between HT and QFT can be eliminated.

## **II. Hole Theory.**

In this section we will derive an expression for the second order change in the vacuum energy due to time dependent perturbation in HT. In order to simplify the discussion and avoid unnecessary mathematical details we will work in 1-1 dimensional space-time where the space dimension is taken along the z-axis and use natural units so that $\hbar = c = 1$. In this case the Dirac equation for a single electron is,

$$i\frac{\partial \psi(z,t)}{\partial t} = (H_0 + qV(z,t))\psi(z,t) \tag{2.1}$$

where,

$$\hat{H}_0 = \left(-i\sigma_x \frac{\partial}{\partial z} + m\sigma_z\right) \tag{2.2}$$

and where 'm' is the mass, q is the electric charge, and $\sigma_x$ and $\sigma_z$ are the Pauli spin matrices,

$$\sigma_x = \begin{pmatrix} 0 & 1 \\ 1 & 0 \end{pmatrix} \quad \text{and} \quad \sigma_z = \begin{pmatrix} 1 & 0 \\ 0 & -1 \end{pmatrix} \tag{2.3}$$

We will assume periodic boundary conditions so that the solutions satisfy

$\psi(z,t) = \psi(z+L,t)$ where L is the 1-dimensional integration volume. In this case the orthonormal free field solutions (V is zero) of (2.1) are given by,

$$\varphi^{(0)}_{\lambda,r}(z,t) = \varphi^{(0)}_{\lambda,r}(z)e^{-i\varepsilon^{(0)}_{\lambda,r}t} = u_{\lambda,r}e^{-i\left(\varepsilon^{(0)}_{\lambda,r}t - p_r z\right)} \qquad (2.4)$$

where 'r' is an integer, $\lambda = \pm 1$ is the sign of the energy, $p_r = 2\pi r/L$, and where,

$$\varepsilon^{(0)}_{\lambda,r} = \lambda E_r \;;\; E_r = \sqrt{p_r^2 + m^2} \;;\; u_{\lambda,r} = N_{\lambda,r}\begin{pmatrix} 1 \\ p_r/(\lambda E_r + m) \end{pmatrix} \;;\; N_{\lambda,r} = \sqrt{\frac{\lambda E_r + m}{2L\lambda E_r}} \qquad (2.5)$$

The quantities $\varphi^{(0)}_{\lambda,r}(z)$ satisfy the relationship,

$$\hat{H}_0 \varphi^{(0)}_{\lambda,r}(z) = \varepsilon^{(0)}_{\lambda,r}\varphi^{(0)}_{\lambda,r}(z) \qquad (2.6)$$

The $\varphi^{(0)}_{\lambda,r}(z)$ form an orthonormal basis set and satisfy,

$$\int_{-L/2}^{+L/2} \varphi^{(0)\dagger}_{\lambda,r}(z)\varphi^{(0)}_{\lambda',s}(z)dz = \delta_{\lambda\lambda'}\delta_{rs} \qquad (2.7)$$

Suppose at some initial time $t_0$ the perturbation V is zero. At this time assume all the negative energy states $\varphi^{(0)}_{-1,r}(z,t_0)$ are occupied by a single electron and all the positive energy states are unoccupied. This is the HT vacuum state. Now let the perturbation V be applied for a period of time and then removed at some final time $t_f$. Therefore V satisfies,

$$V(z,t) = 0 \text{ for } t \le t_0;\; V(z,t) \ne 0 \text{ for } t_0 < t < t_f;\; V(z,t) = 0 \text{ for } t \ge t_f \qquad (2.8)$$

Each initial wave function $\varphi_{\lambda,r}^{(0)}(z,t_0)$ evolves according to the Dirac equation into the final state $\varphi_{\lambda,r}(z,t_f)$. The change in the energy of the state $\varphi_{\lambda,r}^{(0)}(z,t_0)$ from $t_0$ to $t_f$ is given by,

$$\delta\varepsilon_{\lambda,r} = \left\langle \varphi_{\lambda,r}^{\dagger}(z,t_f) H_0 \varphi_{\lambda,r}(z,t_f) \right\rangle - \varepsilon_{\lambda,r}^{(0)} \tag{2.9}$$

where, to simplify notation, we define $\langle g(z) \rangle \equiv \int_{-L/2}^{+L/2} g(z) dz$.

Let $\Delta E_N$ be the sum of the change in energy of the first 2N+1 vacuum electrons, i.e., the sum of the $\delta\varepsilon_{\lambda,r}$ for $\lambda = -1$ and $-N \leq r \leq +N$. Therefore,

$$\Delta E_N = \sum_{r=-N}^{N} \delta\varepsilon_{-1,r} \tag{2.10}$$

The change of the HT vacuum energy, $\Delta E_{hvac}$, is, then, defined as,

$$\Delta E_{hvac} \underset{N \to \infty}{=} \Delta E_N \tag{2.11}$$

(Note that the 'h' in the term $\Delta E_{hvac}$ is to indicate that this is the change in the vacuum energy as determined by HT. This is to distinguish it from the quantity $\Delta E_{qvac}$, to be introduced later, which is the change in the vacuum energy using QFT).

The relationship between the initial and final wave function is,

$$\varphi_{\lambda,r}(z,t_f) = U(t_f,t_0) \varphi_{\lambda,r}^{(0)}(z,t_0) \tag{2.12}$$

where $U(t_f,t_0)$ is a unitary operator. From Thaller [5] a formal expression for $U(t_f,t_0)$ is,

$$U(t_f, t_0) = e^{-iH_0 t_f} \left(1 - iq \int_{t_0}^{t_f} V_I(z,t) dt - q^2 \int_{t_0}^{t_f} V_I(z,t) dt \int_{t_0}^{t} V_I(z,t') dt' + O(q^3)\right) e^{+iH_0 t_0}$$

(2.13)

where $V_I(z,t) = e^{+iH_0 t} V(z,t) e^{-iH_0 t}$. From this we can write,

$$\varphi_{\lambda,r}(z,t_f) = \varphi_{\lambda,r}^{(0)}(z,t_f) + q\varphi_{\lambda,r}^{(1)}(z,t_f) + q^2 \varphi_{\lambda,r}^{(2)}(z,t_f) + O(q^3) \quad (2.14)$$

where $O(q^3)$ means terms to the third order of q or higher. Similarly $\delta\varepsilon_{\lambda,r}$ can be expanded as,

$$\delta\varepsilon_{\lambda,r} = q\delta\varepsilon_{\lambda,r}^{(1)} + q^2 \delta\varepsilon_{\lambda,r}^{(2)} + O(q^3) \quad (2.15)$$

and,

$$\Delta E_N = q\Delta E_N^{(1)} + q^2 \Delta E_N^{(2)} + O(q^3) \quad (2.16)$$

It is shown in Appendix A that the above relationships yield,

$$\delta\varepsilon_{\lambda,r}^{(1)} = 0 \Rightarrow \Delta E_N^{(1)} = 0 \quad (2.17)$$

and,

$$\delta\varepsilon_{\lambda,r}^{(2)} = \sum_{\lambda'=\pm 1} \sum_{s=-\infty}^{+\infty} \left( \left| f_{\lambda',s;\lambda,r} \right|^2 \left( \varepsilon_{\lambda',s}^{(0)} - \varepsilon_{\lambda,r}^{(0)} \right) \right) \quad (2.18)$$

where,

$$f_{\lambda',s;\lambda,r} = \int_{t_0}^{t_f} V_{\lambda',s;\lambda,r}(t) e^{i(\varepsilon_{\lambda',s} - \varepsilon_{\lambda,r})t} dt \quad (2.19)$$

and,

$$V_{\lambda',s;\lambda,r}(t) = \left\langle \varphi_{\lambda',s}^{(0)\dagger}(z) V(z,t) \varphi_{\lambda,r}^{(0)}(z) \right\rangle \quad (2.20)$$

Next use the above results in (2.10) along with (2.15) and (2.16) to obtain,

$$\Delta E_N^{(2)} = \sum_{r=-N}^{N} \left( \sum_{\lambda'=\pm 1} \sum_{s=-\infty}^{+\infty} \left( \left| f_{\lambda',s;-1,r} \right|^2 \left( \varepsilon_{\lambda',s}^{(0)} - \varepsilon_{-1,r}^{(0)} \right) \right) \right) \qquad (2.21)$$

Use (2.5) in the above to yield,

$$\Delta E_N^{(2)} = \sum_{r=-N}^{N} \left( \sum_{\lambda'=\pm 1} \sum_{s=-\infty}^{+\infty} \left( \left| f_{\lambda',s;-1,r} \right|^2 \left( \lambda' E_s + E_r \right) \right) \right) \qquad (2.22)$$

This can be written as,

$$\Delta E_N^{(2)} = \left( Y_N + X_N \right) \qquad (2.23)$$

where,

$$Y_N = \sum_{r=-N}^{N} \left( \sum_{s=-\infty}^{+\infty} \left| f_{+1,s;-1,r} \right|^2 \left( E_s + E_r \right) \right) \qquad (2.24)$$

and

$$X_N = -\sum_{r=-N}^{N} \left( \sum_{s=-\infty}^{+\infty} \left| f_{-1,s;-1,r} \right|^2 \left( E_s - E_r \right) \right) \qquad (2.25)$$

$X_N$ can then be written as,

$$X_N = X_{1,N} + X_{2,N} \qquad (2.26)$$

where,

$$X_{1,N} = -\sum_{r=-N}^{N} \left( \sum_{s=-N}^{+N} \left| f_{-1,s;-1,r} \right|^2 \left( E_s - E_r \right) \right) \qquad (2.27)$$

and,

$$X_{2,N} = -\sum_{r=-N}^{N} \left( \begin{array}{c} \sum_{s=N+1}^{+\infty} \left( \left| f_{-1,s;-1,r} \right|^2 \left( E_s - E_r \right) \right) \\ + \sum_{s=-\infty}^{-N-1} \left( \left| f_{-1,s;-1,r} \right|^2 \left( E_s - E_r \right) \right) \end{array} \right) \qquad (2.28)$$

Now if the dummy indices 's' and 'r' are switched in (2.27) and we use the fact that $\left|f_{-1,s;-1,r}\right|=\left|f_{-1,r;-1,s}\right|$ then we can show that $X_{1,N}=-X_{1,N}=0$ so that,

$$\Delta E_N^{(2)} = \left(X_{2,N}+Y_N\right) \tag{2.29}$$

Therefore the second order change in the HT vacuum energy is,

$$\Delta E_{hvac}^{(2)} \underset{N\to\infty}{=} \Delta E_N^{(2)} = X_{2,N}(N\to\infty)+Y_N(N\to\infty) \tag{2.30}$$

### **III. Quantum field theory.**

In this section we will derive an expression for the change in the vacuum energy due to a time dependent perturbation in QFT. We shall work in the Schrödinger picture. In this case the field operators are time independent and all changes in the system are reflected in the changes of the state vectors. The field operators are defined by,

$$\hat{\psi}(z) = \sum_{\lambda=\pm 1}\sum_{r=-\infty}^{+\infty} \hat{a}_{\lambda,r}\varphi_{\lambda,r}^{(0)}(z); \quad \hat{\psi}^\dagger(z) = \sum_{\lambda=\pm 1}\sum_{r=-\infty}^{+\infty} \hat{a}_{\lambda,r}^\dagger \varphi_{\lambda,r}^{(0)\dagger}(z) \tag{3.1}$$

where the $\hat{a}_{\lambda,r}$ ($\hat{a}_{\lambda,r}^\dagger$) are the destruction(creation) operators for a particle in the state $\varphi_{\lambda,r}^{(0)}(z)$. The operators $\hat{a}_{\lambda,r}$ and $\hat{a}_{\lambda,r}^\dagger$ satisfy the anticommutator relation

$$\hat{a}_{\lambda',s}\hat{a}_{\lambda,r}^\dagger + \hat{a}_{\lambda,r}^\dagger\hat{a}_{\lambda',s} = \delta_{rs}\delta_{\lambda'\lambda}; \text{ all other anticommutators=0} \tag{3.2}$$

The Hamiltonian operator is,

$$\hat{H} = \hat{H}_0 + q\hat{V}(t) \tag{3.3}$$

where,

$$\hat{H}_0 = \int_{-L/2}^{+L/2} \hat{\psi}^\dagger(z) H_0 \hat{\psi}(z) dz - \xi_{ren} = \sum_{\lambda=\pm 1}\sum_{r=-\infty}^{+\infty} \hat{a}_{\lambda,r}^\dagger \hat{a}_{\lambda,r}\varepsilon_{\lambda,r} - \xi_{ren} \tag{3.4}$$

and

$$\hat{V}(t) = \int_{-L/2}^{+L/2} \hat{\psi}^{\dagger}(z) V(z,t) \hat{\psi}(z) dz \tag{3.5}$$

$\xi_{ren}$ is a renormalization constant defined so that the energy of the vacuum state $|0\rangle$ is equal to zero. Following Greiner (see Chapt. 9 of [6]) define the state vector $|0, \text{bare}\rangle$ which is the state vector that is empty of all particles, i.e.,

$$\hat{a}_{\lambda,r} |0, \text{bare}\rangle = 0 \text{ for all } \lambda, r \tag{3.6}$$

The vacuum state vector $|0\rangle$ is defined as the state vector in which all negative energy states are occupied by a single particle. Therefore

$$|0\rangle = \prod_{r=-\infty}^{\infty} \hat{a}^{\dagger}_{-1,r} |0, \text{bare}\rangle \tag{3.7}$$

The vacuum state satisfies the equation,

$$\hat{H}_0 |0\rangle = 0 \tag{3.8}$$

New states are produced by acting on the vacuum state $|0\rangle$ with the operators $\hat{a}_{-1,r}$ and $\hat{a}^{\dagger}_{+1,r}$. The action of the operator $\hat{a}_{-1,r}$ on $|0\rangle$ is to destroy a state with negative energy $-E_r$ and the action of $\hat{a}^{\dagger}_{+1,r}$ is to create a state with positive energy $+E_r$. In either case the energy of the new state is increased. In general if we define some state $|k\rangle$ by,

$$|k\rangle = \hat{a}^{\dagger}_{+1,r_1} \hat{a}^{\dagger}_{+1,r_2} \ldots \hat{a}^{\dagger}_{+1,r_j} \hat{a}_{-1,s_1} \hat{a}_{-1,s_2} \ldots \hat{a}_{-1,s_i} |0\rangle \tag{3.9}$$

then we can write,

$$\hat{H}_0 |k\rangle = \left( \left( E_{r_1} + E_{r_2} + \ldots + E_{r_j} \right) + \left( E_{s_1} + E_{s_2} + \ldots + E_{s_i} \right) \right) |k\rangle \tag{3.10}$$

Now suppose at the initial time $t_0$ the system is in the vacuum state $|0\rangle$. Apply a perturbation per equation (2.8). Under the action of the perturbation the state evolves into the perturbed vacuum state $|0_p(t_f)\rangle$ at the final time $t_f$. The relationship between the final state $|0_p(t_f)\rangle$ and the initial state $|0\rangle$ is given by,

$$|0_p(t_f)\rangle = \hat{U}(t_f, t_0)|0\rangle \tag{3.11}$$

where $\hat{U}(t_f, t_0)$ is a unitary operator. From Sakurai [7] we have,

$$\hat{U}(t_f, t_0) = e^{-i\hat{H}_0 t_f}\left(1 - iq\int_{t_0}^{t_f}\hat{V}_I(t)dt - q^2\int_{t_0}^{t_f}\hat{V}_I(t)dt\int_{t_0}^{t}\hat{V}_I(t')dt' + O(q^3)\right)e^{+i\hat{H}_0 t_0} \tag{3.12}$$

where,

$$\hat{V}_I(t) = e^{i\hat{H}_0 t}\hat{V}(t)e^{-i\hat{H}_0 t} \tag{3.13}$$

The change in the energy of the QFT vacuum state is given by,

$$\Delta E_{qvac} = \langle 0_p(t_f)|\hat{H}_0|0_p(t_f)\rangle - \langle 0|\hat{H}_0|0\rangle \tag{3.14}$$

Use (3.11) and (3.12) in (3.14) along with the fact that $\langle 0|\hat{H}_0 = \hat{H}_0|0\rangle = 0$ to obtain,

$$\Delta E_{qvac} = q^2 \Delta E_{qvac}^{(2)} + O(q^3) \tag{3.15}$$

where,

$$\Delta E_{qvac}^{(2)} = \langle 0|\left(\int_{t_0}^{t_f}\hat{V}_I(t)dt\right)\hat{H}_0\left(\int_{t_0}^{t_f}\hat{V}_I(t)dt\right)|0\rangle \tag{3.16}$$

Use (3.1) in (3.5) to obtain,

$$\hat{V}(t) = \sum_{\lambda=\pm 1}\sum_{\lambda'=\pm 1}\sum_{s=-\infty}^{\infty}\sum_{r=-\infty}^{\infty}\hat{a}^\dagger_{\lambda',s}\hat{a}_{\lambda,r}V_{\lambda',s;\lambda,r}(t) \tag{3.17}$$

Use this result in (3.13) to obtain,

$$\hat{V}_I(t) = \sum_{\lambda=\pm 1}\sum_{\lambda'=\pm 1}\sum_{s=-\infty}^{\infty}\sum_{r=-\infty}^{\infty}\hat{a}^\dagger_{\lambda',s}\hat{a}_{\lambda,r}V_{\lambda',s;\lambda,r}(t)e^{i\left(\varepsilon^{(0)}_{\lambda',s}-\varepsilon^{(0)}_{\lambda,r}\right)t} \quad (3.18)$$

where we have used $e^{i\hat{H}_0 t}\hat{a}^\dagger_{\lambda',s}\hat{a}_{\lambda,r}e^{-i\hat{H}_0 t} = \hat{a}^\dagger_{\lambda',s}\hat{a}_{\lambda,r}e^{+i\left(\varepsilon^{(0)}_{\lambda',s}-\varepsilon^{(0)}_{\lambda,r}\right)t}$. From this we obtain,

$$\int_{t_0}^{t_f}\hat{V}_I(t)dt = \sum_{\lambda=\pm 1}\sum_{\lambda'=\pm 1}\sum_{s=-\infty}^{\infty}\sum_{r=-\infty}^{\infty}\hat{a}^\dagger_{\lambda',s}\hat{a}_{\lambda,r}f_{\lambda',s;\lambda,r} \quad (3.19)$$

Use this in (3.16) to yield,

$$\Delta E^{(2)}_{qvac} = \sum_{\lambda,\lambda',\eta,\eta'=\pm 1}\sum_{j,k,r,s=-\infty}^{\infty}\langle 0|\hat{a}^\dagger_{\eta',j}\hat{a}_{\eta,k}\hat{H}_0\hat{a}^\dagger_{\lambda',s}\hat{a}_{\lambda,r}|0\rangle\left(f_{\eta',j;\eta,k}\right)\left(f_{\lambda',s;\lambda,r}\right) \quad (3.20)$$

From (3.7), (3.1), and (3.10) we have the relationship,

$$\langle 0|\hat{a}^\dagger_{\eta',j}\hat{a}_{\eta,k}\hat{H}_0\hat{a}^\dagger_{\lambda',s}\hat{a}_{\lambda,r}|0\rangle = \delta_{-1,\lambda}\delta_{+1,\lambda'}\left(\varepsilon_{\lambda',s}-\varepsilon_{\lambda,r}\right)\delta_{rj}\delta_{sk}\delta_{\lambda\eta'}\delta_{\lambda'\eta} \quad (3.21)$$

Note that this quantity is zero if $\lambda = +1$ or $\lambda' = -1$. Use this in (3.20) to obtain,

$$\Delta E^{(2)}_{qvac} = \sum_{r=-\infty}^{\infty}\sum_{s=-\infty}^{\infty}\left(\varepsilon_{+1,s}-\varepsilon_{-1,r}\right)\left|\left(f_{-1,r;+1,s}\right)\right|^2 \quad (3.22)$$

where we have used $\left|\left(f_{-1,r;+1,s}\right)\right|^2 = \left(f_{-1,r;+1,s}\right)\left(f_{+1,s;-1,r}\right)$. Use (2.5) in the above to obtain,

$$\Delta E^{(2)}_{qvac} = \sum_{r=-\infty}^{\infty}\sum_{s=-\infty}^{\infty}\left(E_s+E_r\right)\left|\left(f_{-1,r;+1,s}\right)\right|^2 \quad (3.23)$$

Now compare this with $\Delta E^{(2)}_{hvac}$ of equation (2.30). If we examine (2.24) we see that $\Delta E^{(2)}_{qvac}$ appears to be equal to $Y_N(N\to\infty)$ (Replace N with $\infty$ in equation (2.24)). Therefore for $\Delta E^{(2)}_{qvac}$ to be equal to $\Delta E^{(2)}_{hvac}$ we must have that $X_{2,N}(N\to\infty)$ is equal to zero. So the question that must be determined is does $X_{2,N}(N\to\infty)$ equal zero for all possible perturbations. It will be shown in the next section that this in not the case.

## IV. Calculating the change in energy.

In the previous sections we have derived formal relationships for the second order change in the vacuum energy of HT and QFT. It has been shown that the formal expressions are not identical. In this section we shall work an actual problem which will clearly demonstrate the differences between HT and QFT. Let the perturbation $V(z,t)$ be given by,

$$V(z,t) = 4\cos(k_w z)\left(\frac{\sin(mt)}{t}\right) = \left(e^{ik_w z} + e^{-ik_w z}\right)\int_{-m}^{+m} e^{iqt} dq \qquad (4.1)$$

where m is the mass of the electron and $k_w = 2\pi w/L < m$ where w is a positive integer. It is obvious from the above expression that $V(z,t) \to 0$ at $t \to \pm\infty$. Under the action of this electric potential each initial wave function $\varphi_{\lambda,r}^{(0)}(z,t_0)$, where $t_0 \to -\infty$, evolves into the final wave function $\varphi_{\lambda,r}(z,t_f)$ where $t_f \to +\infty$. Use (4.1) in (2.19) to obtain,

$$f_{\lambda',s;\lambda,r} = \left\langle \varphi_{\lambda',s}^{(0)\dagger}(z)\left(e^{ik_w z} + e^{-ik_w z}\right)\varphi_{\lambda,r}^{(0)}(z) \right\rangle \int_{-\infty}^{+\infty} e^{i(\varepsilon_{\lambda',s} - \varepsilon_{\lambda,r})t}\left(\int_{-m}^{+m} e^{iqt} dq\right) dt \qquad (4.2)$$

Use (2.4) in the above to obtain,

$$f_{\lambda',s;\lambda,r} = u_{\lambda',s}^{\dagger} u_{\lambda,r} \int_{-\infty}^{+\infty} dt \int_{-L/2}^{+L/2} dz \begin{pmatrix} e^{-i(\lambda E_r - \lambda' E_s)t} e^{i(p_r - p_s)z} \\ \times \left(e^{ik_w z} + e^{-ik_w z}\right)\int_{-m}^{+m} e^{iqt} dq \end{pmatrix} \qquad (4.3)$$

Perform the integrations over t and z to obtain,

$$f_{\lambda',s;\lambda,r} = 2\pi\left\{ u_{\lambda',s}^{\dagger} u_{\lambda,r} \begin{pmatrix} \delta_L(w+r-s) \\ +\delta_L(-w+r-s) \end{pmatrix} \int_{-m}^{+m} \delta\begin{pmatrix} -\lambda E_r \\ +\lambda' E_s + q \end{pmatrix} dq \right\} \qquad (4.4)$$

where,

$$\delta_L(r) = \begin{cases} L & \text{if } r = 0 \\ 0 & \text{if } r \neq 0 \end{cases} \tag{4.5}$$

From the definition of the delta function we have the following relationship

$$\delta_L(\pm w + r - s) \int_{-m}^{+m} \delta\begin{pmatrix} -\lambda E_r \\ +\lambda' E_s + q \end{pmatrix} dq = \delta_L(\pm w + r - s) \int_{-m}^{+m} \delta\begin{pmatrix} -\lambda E_{s \mp w} \\ +\lambda' E_s + q \end{pmatrix} dq \tag{4.6}$$

Due to the fact that $(E_r + E_s) \geq 2m$ and $k_w < m$ we have that,

$$\int_{-m}^{+m} \delta(-\lambda E_{s \mp w} + \lambda' E_s + q) dq = \delta_{\lambda \lambda'} \tag{4.7}$$

Use these results in (4.4) to obtain,

$$f_{\lambda', s; \lambda, r} = 2\pi \begin{pmatrix} u^\dagger_{\lambda', r+w} u_{\lambda, r} \delta_L(w + r - s) \\ + u^\dagger_{\lambda', r-w} u_{\lambda, r} \delta_L(-w + r - s) \end{pmatrix} \delta_{\lambda \lambda'} \tag{4.8}$$

This yields,

$$|f_{\lambda', s; \lambda, r}|^2 = 4\pi^2 L \begin{pmatrix} |u^\dagger_{\lambda', r+w} u_{\lambda, r}|^2 \delta_L(w + r - s) \\ + |u^\dagger_{\lambda', r-w} u_{\lambda, r}|^2 \delta_L(-w + r - s) \end{pmatrix} \delta_{\lambda \lambda'} \tag{4.9}$$

where we have used $(\delta_{\lambda \lambda'})^2 = \delta_{\lambda \lambda'}$, $(\delta_L(r))^2 = L \delta_L(r)$, and

$\delta_L(w + r - s) \delta_L(-w + r - s) = 0$. Use (4.9) in (2.18) to obtain,

$$\delta \varepsilon^{(2)}_{\lambda, r} = 4\pi^2 L^2 \begin{pmatrix} |u^\dagger_{\lambda, r+w} u_{\lambda, r}|^2 \left( \varepsilon^{(0)}_{\lambda, r+w} - \varepsilon^{(0)}_{\lambda, r} \right) \\ + |u^\dagger_{\lambda, r-w} u_{\lambda, r}|^2 \left( \varepsilon^{(0)}_{\lambda, r-w} - \varepsilon^{(0)}_{\lambda, r} \right) \end{pmatrix} \tag{4.10}$$

It is shown in Appendix B that,

$$\delta \varepsilon^{(2)}_{\lambda, r} = 2\pi^2 \lambda k_w \left( \frac{(p_r + k_w)}{E_{r+w}} - \frac{(p_r - k_w)}{E_{r-w}} \right) \tag{4.11}$$

Now, in order to evaluate we $\Delta E^{(2)}_{hvac}$ are interested in the quantity $\delta\varepsilon^{(2)}_{\lambda,r}$ for the negative energy states in which case $\lambda = -1$. Therefore, for negative energy states,

$$\delta\varepsilon^{(2)}_{-1,r} = -2\pi^2 k_w \left( \frac{(p_r + k_w)}{E_{r+w}} - \frac{(p_r - k_w)}{E_{r-w}} \right) \tag{4.12}$$

It is shown in Appendix C that this quantity is negative for all $\delta\varepsilon^{(2)}_{-1,r}$. When this result is used in (2.10) and (2.11) it is evident that $\Delta E^{(2)}_{hvac} < 0$. Now use the above results to evaluate $\Delta E^{(2)}_{qvac}$ (see equation (3.23)). From (4.9) we have that $f_{-1,r;+1,s} = 0$ because $\lambda \neq \lambda'$. Therefore $\Delta E^{(2)}_{qvac} = 0$ so that $\Delta E^{(2)}_{hvac} \neq \Delta E^{(2)}_{qvac}$. Now what accounts for the difference between HT and QFT. Recall that $\Delta E^{(2)}_{hvac}$ (see equation (2.30)) consists of two terms $Y_N(N \to \infty)$ and $X_{2,N}(N \to \infty)$. Due to the fact that $f_{-1,r;+1,s} = 0$ we have that $Y_N(N \to \infty) = 0$. Therefore the difference between $\Delta E^{(2)}_{qvac}$ and $\Delta E^{(2)}_{hvac}$ is due to the term $X_{2,N}(N \to \infty)$ which must be negative even as $N \to \infty$.

### V. Evaluating $X_{2,N}$

In this section we will calculate $X_{2,N}(N \to \infty)$. Refer to (2.28) to obtain,

$$X_{2,N} = X_{2A,N} + X_{2B,N} \tag{5.1}$$

where,

$$X_{2A,N} = -\sum_{r=-N}^{N} \sum_{s=N+1}^{+\infty} \left( \left| f_{-1,s;-1,r} \right|^2 (E_s - E_r) \right) \tag{5.2}$$

and

$$X_{2B,N} = -\sum_{r=-N}^{N} \sum_{s=-\infty}^{-N-1} \left( \left| f_{-1,s;-1,r} \right|^2 (E_s - E_r) \right) \tag{5.3}$$

To evaluate $X_{2A,N}$ use (4.9) in (5.2) to obtain,

$$X_{2A,N} = -4\pi^2 L \sum_{r=-N}^{N} \sum_{s=N+1}^{+\infty} \left( \left( \begin{array}{l} \left| u^\dagger_{-1,r+w} u_{-1,r} \right|^2 \delta_L(w+r-s) \\ + \left| u^\dagger_{-1,r-w} u_{-1,r} \right|^2 \delta_L(-w+r-s) \end{array} \right) (E_s - E_r) \right) \quad (5.4)$$

Now we will evaluate this expression for $N \to \infty$. In the above expression $r-s < 0$ so that for positive w, $\delta_L(-w+r-s) = 0$ and $\delta_L(w+r-s) = L$ if $s-r = w$. From this we obtain,

$$X_{2A,N}(N \to \infty) = -4\pi^2 L^2 \sum_{r=N+1-w}^{N} \left( \left| u^\dagger_{-1,r+w} u_{-1,r} \right|^2 (E_{r+w} - E_r) \right) \quad (5.5)$$

From (B.3) we have,

$$X_{2A,N}(N \to \infty) = -2\pi^2 \sum_{r=N+1-w}^{N} \left( \left( 1 + \frac{p_r(p_r + k_w) + m^2}{(E_{r+w})(E_r)} \right) (E_{r+w} - E_r) \right) \quad (5.6)$$

Now for large 'r',

$$(E_{r+w} - E_r) = \sqrt{(p_r + k_w)^2 + m^2} - \sqrt{(p_r)^2 + m^2} \underset{r \to \infty}{=} k_w \quad (5.7)$$

and,

$$\frac{p_r(p_r + k_w) + m^2}{(E_{r+w})(E_r)} \underset{r \to \infty}{=} 1 \quad (5.8)$$

Use this in (5.6) to obtain,

$$X_{2A,N}(N \to \infty) = -2\pi^2 \sum_{r=N+1-w}^{N} 2k_w = -2\pi^2(2wk_w) = -2\pi L(k_w)^2 \quad (5.9)$$

Apply the same reasoning to $X_{2B,N}(N \to \infty)$ to obtain,

$$X_{2B,N}(N \to \infty) = -2\pi L(k_w)^2 \quad (5.10)$$

Use these results in (5.1) to obtain,

$$X_{2,N}(N \to \infty) = -4\pi L(k_w)^2 \tag{5.11}$$

Use this in (2.30) along with the fact that $Y_N(N \to \infty) = 0$ to obtain,

$$\Delta E^{(2)}_{hvac} = -4\pi L(k_w)^2 \tag{5.12}$$

### **VI. Redefining the QFT vacuum state.**

The reason there is a difference between HT and QFT is due the presence of the term $X_{2,N}(N \to \infty)$ in the expression for the second order change in the HT vacuum energy, $\Delta E^{(2)}_{hvac}$. This term does not appear in the expression for the change of the QFT vacuum energy, $\Delta E^{(2)}_{qvac}$. This is similar to the situation that exists when a time independent perturbation is applied as discussed in [4]. In this case it was shown that HT and QFT could be made equivalent by redefining the QFT vacuum state $|0\rangle$. We will show that this result also applies to the case of the time dependent potential which has been discussed here. Redefine the QFT vacuum state as follows:

$$|0_N\rangle \underset{N \to \infty}{=} \prod_{r=-N}^{N} \hat{a}^{\dagger}_{-1,r} |0, \text{bare}\rangle \tag{6.1}$$

According to this definition the band of vacuum states with energies between $-m$ and $-E_N$ are occupied by a single electron. All positive energy states are unoccupied. All states with energy less than $-E_N$ are also unoccupied. This differs from the standard definition of the vacuum state $|0\rangle$ given by equation (3.7). The difference is due to the fact that the lower edge of the negative energy band is defined using a limiting procedure.

4New states are produced by acting on the vacuum state $|0_N\rangle$ with $\hat{a}^{\dagger}_{+1,r}$ for all r, $\hat{a}_{-1,r}$ where $-E_r > -E_N$, and $\hat{a}^{\dagger}_{-1,r}$ where $-E_N > -E_r$. The action of the operator $\hat{a}_{-1,r}$ or $\hat{a}^{\dagger}_{+1,r}$ on $|0_N\rangle$ is to increase the energy. The action of the operator $\hat{a}^{\dagger}_{-1,r}$ where $-E_N > -E_r$ is to add a state with energy $-E_r$ underneath the occupied band with decreases the energy of the new state. (For additional discussion of these redefined vacuum state see references [4] and [8]).

Referring to equation (3.16) and the discussion in Section III we obtain,

$$\Delta E^{(2)}_{Nqvac} = \langle 0_N | \left( \int_{t_0}^{t_f} \hat{V}_I(t)\,dt \right) \hat{H}_0 \left( \int_{t_0}^{t_f} \hat{V}_I(t)\,dt \right) | 0_N \rangle \tag{6.2}$$

where $\Delta E^{(2)}_{Nqvac}$ is the second order change in the energy of the redefined vacuum state $|0_N\rangle$. From this and the discussions leading up to (3.20) we obtain,

$$\Delta E^{(2)}_{Nqvac} = Y'_N(N \to \infty) + X'_{2,N}(N \to \infty) \tag{6.3}$$

where,

$$Y'_N = \sum_{r=-N}^{N} \sum_{s=-\infty}^{\infty} \langle 0_N | \hat{a}^{\dagger}_{-1,r}\hat{a}_{+1,s}\hat{H}_0\hat{a}^{\dagger}_{+1,s}\hat{a}_{-1,r} | 0_N \rangle \left| \left( f_{-1,r;+1,s} \right) \right|^2 \tag{6.4}$$

and,

$$X'_{2,N} = \sum_{r=-N}^{N} \left\{ \sum_{s=N+1}^{\infty} \langle 0_N | \hat{a}^{\dagger}_{-1,r}\hat{a}_{-1,s}\hat{H}_0\hat{a}^{\dagger}_{-1,s}\hat{a}_{-1,r} | 0_N \rangle \left| \left( f_{-1,r;-1,s} \right) \right|^2 \right. \\ \left. + \sum_{s=-\infty}^{-N-1} \langle 0_N | \hat{a}^{\dagger}_{-1,r}\hat{a}_{-1,s}\hat{H}_0\hat{a}^{\dagger}_{-1,s}\hat{a}_{-1,r} | 0_N \rangle \left| \left( f_{-1,r;-1,s} \right) \right|^2 \right\} \tag{6.5}$$

The above can be evaluated to obtain,

$$Y'_N = \sum_{r=-N}^{N} \sum_{s=-\infty}^{\infty} \left| (E_s + E_r)\left( f_{-1,r;+1,s} \right) \right|^2 \tag{6.6}$$

and,

$$X'_{2,N} = -\sum_{r=-N}^{N} \left\{ \begin{array}{l} \sum_{s=N+1}^{\infty} (E_s - E_r)\left|(f_{-1,r;-1,s})\right|^2 \\ + \sum_{s=-\infty}^{-N-1} (E_s - E_r)\left|(f_{-1,r;-1,s})\right|^2 \end{array} \right\} \quad (6.7)$$

Compare these equations with (2.24) and (2.28) to show that $Y'_N = Y_N$ and $X'_{2,N} = X_{2,N}$. Therefore hole theory and field theory yield identical results when the vacuum state is defined per equation (6.1). This result is consistent with the result obtained in the case of the time independent perturbation as discussed in [4].

## VII. Conclusion

We have derived expressions for the second order change of the vacuum energy due to a time dependent perturbation using HT and QFT. The formal result for HT is given by (2.30) and the result for QFT is given by (3.23). There is a difference in these results due to the presence of the term $X_{2,N}(N \to \infty)$ which appears in the result in HT but not in QFT. In order to understand the importance of the term $X_{2,N}(N \to \infty)$ we have worked on actual problem which is presented in Section III. It was shown for this problem that the change in energy of every negative energy electron is negative. Therefore $E^{(2)}_{hvac} < 0$. However it is shown for this same problem that $E^{(2)}_{qvac} = 0$. The difference between $E^{(2)}_{hvac}$ and $E^{(2)}_{qvac}$ is the term $X_{2,N}(N \to \infty)$. It was shown in Section V that this term is nonzero. If the definition of the QFT vacuum is redefined according to the discussion in Section VI then the term $X_{2,N}(N \to \infty)$ appears in the QFT result and HT and QFT give equivalent results.

## Appendix A

Since the Dirac equation does not affect the normalization condition we have,

$$\left\langle \varphi_{\lambda,r}^{\dagger}(z,t_f)\varphi_{\lambda,r}(z,t_f)\right\rangle = \left\langle \varphi_{\lambda,r}^{(0)\dagger}(z,t_f)\varphi_{\lambda,r}^{(0)}(z,t_f)\right\rangle = 1 \tag{A.1}$$

Use (2.14) in the above to obtain,

$$0 = q\left(\left\langle \varphi_{\lambda,r}^{(0)\dagger}\varphi_{\lambda,r}^{(1)}\right\rangle + \left\langle \varphi_{\lambda,r}^{(1)\dagger}\varphi_{\lambda,r}^{(0)}\right\rangle\right) + q^2\left(\left\langle \varphi_{\lambda,r}^{(0)\dagger}\varphi_{\lambda,r}^{(2)}\right\rangle + \left\langle \varphi_{\lambda,r}^{(2)\dagger}\varphi_{\lambda,r}^{(0)}\right\rangle + \left\langle \varphi_{\lambda,r}^{(1)\dagger}\varphi_{\lambda,r}^{(1)}\right\rangle\right) \tag{A.2}$$

This yields,

$$q^2\left\langle \varphi_{\lambda,r}^{(1)\dagger}\varphi_{\lambda,r}^{(1)}\right\rangle = -q\left(\left\langle \varphi_{\lambda,r}^{(0)\dagger}\varphi_{\lambda,r}^{(1)}\right\rangle + \left\langle \varphi_{\lambda,r}^{(1)\dagger}\varphi_{\lambda,r}^{(0)}\right\rangle\right) - q^2\left(\left\langle \varphi_{\lambda,r}^{(0)\dagger}\varphi_{\lambda,r}^{(2)}\right\rangle + \left\langle \varphi_{\lambda,r}^{(2)\dagger}\varphi_{\lambda,r}^{(0)}\right\rangle\right) \tag{A.3}$$

Next use (2.14) in (2.15) to obtain,

$$\delta\varepsilon_{\lambda,r} = \left\langle \varphi_{\lambda,r}^{(0)\dagger}H_0\varphi_{\lambda,r}^{(0)}\right\rangle + q\begin{pmatrix}\left\langle \varphi_{\lambda,r}^{(1)\dagger}H_0\varphi_{\lambda,r}^{(0)}\right\rangle \\ +\left\langle \varphi_{\lambda,r}^{(0)\dagger}H_0\varphi_{\lambda,r}^{(1)}\right\rangle\end{pmatrix} + q^2\begin{pmatrix}\left\langle \varphi_{\lambda,r}^{(2)\dagger}H_0\varphi_{\lambda,r}^{(0)}\right\rangle + \left\langle \varphi_{\lambda,r}^{(1)\dagger}H_0\varphi_{\lambda,r}^{(1)}\right\rangle \\ +\left\langle \varphi_{\lambda,r}^{(0)\dagger}H_0\varphi_{\lambda,r}^{(2)}\right\rangle\end{pmatrix} + O(q^3) - \varepsilon_{\lambda,r}^{(0)}$$

$$\tag{A.4}$$

Use (2.6) in the above to obtain,

$$\delta\varepsilon_{\lambda,r} = q\varepsilon_{\lambda,r}^{(0)}\begin{pmatrix}\left\langle \varphi_{\lambda,r}^{(1)\dagger}\varphi_{\lambda,r}^{(0)}\right\rangle \\ +\left\langle \varphi_{\lambda,r}^{(0)\dagger}\varphi_{\lambda,r}^{(1)}\right\rangle\end{pmatrix} + q^2\varepsilon_{\lambda,r}^{(0)}\begin{pmatrix}\left\langle \varphi_{\lambda,r}^{(2)\dagger}\varphi_{\lambda,r}^{(0)}\right\rangle \\ +\left\langle \varphi_{\lambda,r}^{(0)\dagger}\varphi_{\lambda,r}^{(2)}\right\rangle\end{pmatrix} + q^2\left\langle \varphi_{\lambda,r}^{(1)\dagger}H_0\varphi_{\lambda,r}^{(1)}\right\rangle + O(q^3) \tag{A.5}$$

Use (A.3) in the above to obtain,

$$\delta\varepsilon_{\lambda,r} = q^2\left\langle \varphi_{\lambda,r}^{(1)\dagger}H_0\varphi_{\lambda,r}^{(1)}\right\rangle - q^2\varepsilon_{\lambda,r}^{(0)}\left\langle \varphi_{\lambda,r}^{(1)\dagger}\varphi_{\lambda,r}^{(1)}\right\rangle + O(q^3) \tag{A.6}$$

Therefore,

$$\delta\varepsilon_{\lambda,r}^{(1)} = 0 \Rightarrow \Delta E_N^{(1)} = 0 \tag{A.7}$$

and

$$\delta\varepsilon_{\lambda,r}^{(2)} = \left\langle \varphi_{\lambda,r}^{(1)\dagger}(z,t_f) H_0 \varphi_{\lambda,r}^{(1)}(z,t_f) \right\rangle - \varepsilon_{\lambda,r}^{(0)} \left\langle \varphi_{\lambda,r}^{(1)\dagger}(z,t_f) \varphi_{\lambda,r}^{(1)}(z,t_f) \right\rangle \quad (A.8)$$

Evaluate this equation as follows. From (2.12), (2.13), and (2.14) we obtain,

$$\varphi_{\lambda,r}^{(1)}(z,t_f) = -i \int_{t_0}^{t_f} e^{-iH_0(t_f-t)} V(z,t) e^{-iH_0(t-t_0)} \varphi_{\lambda,r}^{(0)}(z,t_0) dt \quad (A.9)$$

This becomes,

$$\varphi_{\lambda,r}^{(1)}(z,t_f) = -i \int_{t_0}^{t_f} e^{-iH_0(t_f-t)} V(z,t) \varphi_{\lambda,r}^{(0)}(z,t) dt \quad (A.10)$$

Define,

$$V_{\lambda',s;\lambda,r}(t) = \left\langle \varphi_{\lambda',s}^{(0)\dagger}(z) V(z,t) \varphi_{\lambda,r}^{(0)}(z) \right\rangle \quad (A.11)$$

Since the $\varphi_{\lambda,r}^{(0)}(z)$ form on orthonormal basis we can expand the quantity $V(z,t)\varphi_{\lambda,r}^{(0)}(z)$ as a Fourier expansion as follows,

$$V(z,t)\varphi_{\lambda,r}^{(0)}(z) = \sum_{\lambda'=\pm 1} \sum_{s=-\infty}^{+\infty} \varphi_{\lambda',s}^{(0)}(z) V_{\lambda',s;\lambda,r}(t) \quad (A.12)$$

Use this in (A.10) to obtain,

$$\varphi_{\lambda,r}^{(1)}(z,t_f) = -i \sum_{\lambda'=\pm 1} \sum_{s=-\infty}^{+\infty} \varphi_{\lambda',s}^{(0)}(z,t_f) f_{\lambda',s;\lambda,r} \quad (A.13)$$

where,

$$f_{\lambda',s;\lambda,r} = \int_{t_0}^{t_f} V_{\lambda',s;\lambda,r}(t) e^{i\left(\varepsilon_{\lambda',s}^{(0)} - \varepsilon_{\lambda,r}^{(0)}\right)t} dt \quad (A.14)$$

Use this in (A.8) to obtain,

$$\delta\varepsilon_{\lambda,r}^{(2)} = \sum_{\lambda'=\pm 1} \sum_{s=-\infty}^{+\infty} \left( \left| f_{\lambda',s;\lambda,r} \right|^2 \left( \varepsilon_{\lambda',s}^{(0)} - \varepsilon_{\lambda,r}^{(0)} \right) \right) \quad (A.15)$$

## Appendix B

We want to evaluate equation (4.10) and show that it results in equation (4.11). Use (2.5) to obtain,

$$u^\dagger_{\lambda,r\pm w} u_{\lambda,r} = \sqrt{\frac{\lambda E_{r\pm w} + m}{2\lambda L E_{r\pm w}}} \sqrt{\frac{\lambda E_r + m}{2\lambda L E_r}} \left(1 + \frac{p_r(p_r + k)}{(\lambda E_{r\pm w} + m)(\lambda E_r + m)}\right) \quad (B.1)$$

Use $(\lambda E_r)^2 - m^2 = p_r^2$ in the above to obtain,

$$u^\dagger_{\lambda,r\pm w} u_{\lambda,r} = \sqrt{\frac{\lambda E_{r\pm w} + m}{2\lambda L E_{r\pm w}}} \sqrt{\frac{\lambda E_r + m}{2\lambda L E_r}} \left(1 + \frac{(\lambda E_{r\pm w} - m)(\lambda E_r - m)}{p_r(p_r \pm k_w)}\right) \quad (B.2)$$

Use this result to yield,

$$\left|u^\dagger_{\lambda,r\pm w} u_{\lambda,r}\right|^2 = \left(\frac{(\lambda E_{r\pm w})(\lambda E_r) + p_r(p_r \pm k_w) + m^2}{2(\lambda E_{r\pm w})(\lambda E_r) L^2}\right) \quad (B.3)$$

Use this in (4.10) along with the fact that $\lambda^2 = 1$ to obtain

$$\delta\varepsilon^{(2)}_{\lambda,r} = 2\pi^2\lambda \left(\begin{array}{l}\left(1 + \dfrac{p_r(p_r + k_w) + m^2}{E_{r+w} E_r}\right)(E_{r+w} - E_r) \\ + \left(1 + \dfrac{p_r(p_r - k_w) + m^2}{E_{r-w} E_r}\right)(E_{r-w} - E_r)\end{array}\right) \quad (B.4)$$

This yields,

$$\delta\varepsilon^{(2)}_{\lambda,r} = 2\pi^2\lambda \left(\begin{array}{l}\left(1 + \dfrac{E_{r+w}}{E_r} - \dfrac{k(p_r + k_w)}{E_{r+w} E_r}\right)(E_{r+w} - E_r) \\ +(w \to -w)\end{array}\right) \quad (B.5)$$

Some additional algebraic manipulation yields,

$$\delta\varepsilon^{(2)}_{\lambda,r} = 2\pi^2\lambda \left(\begin{array}{l}\left(-E_r + \dfrac{E^2_{r+w}}{E_r} - \dfrac{k_w(p_r + k_w)}{E_r} + \dfrac{k_w(p_r + k_w)}{E_{r+w}}\right) \\ +(w \to -w)\end{array}\right) \quad (B.6)$$

Use some simple algebra to obtain,

$$\delta\varepsilon^{(2)}_{\lambda,r} = 2\pi^2\lambda\left(\left(\frac{p_r k_w}{E_r} + \frac{k_w(p_r+k_w)}{E_{r+w}}\right) + \left(\frac{-p_r k_w}{E_r} - \frac{k_w(p_r-k_w)}{E_{r-w}}\right)\right) \quad (B.7)$$

Use this result to yield (4.11).

## Appendix C

Assume $k_w$ is positive. Then it can be shown that,

$$\frac{(p_r+k_w)}{E_{r+w}} > \frac{(p_r-k_w)}{E_{r-w}} \quad \text{for all } p_r \quad (C.1)$$

First consider the case where $p_r$ is positive. The relationship is obviously true for $k_w > p_r$. Now let $p_r > k_w$. In this case both sides of (C.1) are positive therefore we can square both sides to obtain,

$$(p_r+k_w)^2 E^2_{r-w} > (p_r-k_w)^2 E^2_{r+w} \quad (C.2)$$

From this we obtain,

$$(p_r+k_w)^2\left((p_r-k_w)^2+m^2\right) > (p_r-k_w)^2\left((p_r+k_w)^2+m^2\right) \quad (C.3)$$

This yields,

$$(p_r+k_w)^2 > (p_r-k_w)^2 \quad (C.4)$$

which is true for positive $k_w$ and $p_r > k_w$. If $p_r$ is negative then (C.1) becomes,

$$\frac{(-|p_r|+k_w)}{E_{|r|-w}} > \frac{(-|p_r|-k_w)}{E_{|r|+w}} \quad (C.5)$$

This yields,

$$\frac{(|p_r|+k_w)}{E_{|r|+w}} > \frac{(|p_r|-k_w)}{E_{|r|-w}} \qquad (C.6)$$

which is obviously true from the previous discussion.

### References

1. F.A.B. Coutinho, D. Kaing, Y. Nagami, and L. Tomio, Can. J. of Phys., **80**, 837 (2002). (see also quant-ph/0010039).

2. F.A.B. Coutinho, Y. Nagami, and L. Tomio, Phy. Rev. A, **59**, 2624 (1999).

3. R. M. Cavalcanti, quant-ph/9908088.

4. D. Solomon. Can. J. Phys., **81**, 1165, (2003).

5. B. Thaller, "The Dirac Equation", Springer-Verlag, Berlin (1992).

6. W. Greiner, B. Muller, and T. Rafelski, "Quantum Electrodynamics of Strong Fields", Springer-Verlag, Berlin (1985).

7. J.J. Sakurai, "Advanced Quantum Mechanics", Addison-Wesley Publishing Company, Inc., Redwood City, California, (1967).

8. D. Solomon, Can. J. Phys. **76**, 111 (1998). (see also quant-ph/9905021).